\documentclass[useAMS,usenatbib,usegraphicx]{mn2e}
\usepackage{amssymb}
\usepackage{graphicx}
\usepackage{color}

\title[Photometric study of the Blazhko star {DM~Cyg}]{An extensive 
photometric study of the Blazhko RR Lyrae star DM~Cyg\thanks{Based on observations collected with the automatic 60 cm telescope of Konkoly Observatory, Budapest, Sv\'abhegy}}
\author[J. Jurcsik et al.]{J. Jurcsik$^{1}$
Zs. Hurta$^{1,2}$,  \'A.  S\'odor$^{1}$, B. Szeidl$^{1}$, I. Nagy$^{2}$,  K. Posztob\'anyi$^{4}$,  \and  M. V\'aradi$^{1,3}$, K. Vida$^{1,2}$, B. Belucz$^{2}$, I. D\'ek\'any$^{1}$, G. Hajdu$^{2}$, Zs. K\H ov\'ari$^{1}$, E. Kun$^{5}$ \\
\\
$^{1}$Konkoly Observatory of the Hungarian Academy of Sciences, H-1525 Budapest PO Box 67, Hungary\\
$^{2}$E\"otv\"os University, Dept. of Astronomy, H-1518 Budapest PO Box 49, Hungary\\
$^{3}$Observatoire de Geneve, Universite de Gen\`eve, CH-1290, Sauverny, Switzerland\\
$^{4}$AEKI, KFKI Atomic Energy Research Institute, Thermohydraulic Department, H-1525 Budapest 114, PO Box 49, Hungary\\
$^{5}$Department of Experimental Physics and Astronomical Observatory, University of Szeged, 6720 Szeged, D\'om t\'er 9, Hungary}
\begin{document}

\date{Accepted 2009 December 15. Received 2009 December 14; in original form 2009 January 22}

\pagerange{\pageref{firstpage}--\pageref{lastpage}} \pubyear{2009}

\maketitle

\label{firstpage}

\begin{abstract}

DM Cyg, a fundamental mode RRab star was observed in the 2007 and 2008 seasons in the frame of the Konkoly Blazhko Survey. Very small amplitude light curve modulation was detected with 10.57 d modulation period. The maximum brightness and phase variations do not exceed 0.07 mag and 7 min, respectively. In spite of the very small amplitude of the modulation, beside the frequency triplets characterizing the Fourier spectrum of the light curve two quintuplet components were also identified. The accuracy and the good phase coverage of our observations made it possible to analyse the light curves at different phases of the modulation separately. Utilizing the IP method (S\'odor, Jurcsik and Szeidl, 2009) we could detect very small systematic changes in the global mean physical parameters of DM Cyg during its Blazhko cycle. The detected changes are similar to what we have already found for a large modulation amplitude Blazhko variable MW Lyrae. The amplitudes of the detected changes in the physical parameters  of DM Cyg are only about 10\% of that what have been found in MW Lyr. This is in accordance with its small modulation amplitude being about one tenth of the modulation amplitude of MW Lyr.

The pulsation period of DM Cyg has been increasing by a rate of $\beta = 0.091 \mathrm{ d Myr^{-1}}$ during the hundred-year time base of the observations. Konkoly archive photographic observations indicate that when the pulsation period of the variable was shorter by $\Delta p_{\mathrm puls} = 5 \cdot 10^{-6}$ d the modulation period was longer by $\Delta p_{\mathrm mod} = 0.066$ d than today.

\end{abstract}

\begin{keywords}
stars: variables: other -- 
stars: horizontal branch --
stars: individual: DM~Cyg -- 
techniques: photometric -- 
methods: data analysis -- 
\end{keywords}

\section{Introduction}

Utilizing our full access to an automatic 60 cm telescope we have obtained extended multicolour observations of many fundamental mode RR Lyrae variables showing light curve modulation (the Blazhko effect) during the past five years. Detailed analyses of some of our targets were already published in \cite{rrgI,ssc,mw1,mw2}. Our observations are the first multicolour photometric data which are accurate and dense enough to allow not only to determine the frequencies appearing in the Fourier spectra of the light curves, but also to find changes in the global mean physical parameters of the stars ($L, T_{eff}, R$) in different phases of the modulation. In order to extract changes in the physical parameters exclusively from multicolour photometric data we developed an inverse photometric Baade-Wesselink method  \cite[IPM;][]{ipm}. The IPM was successfully applied to determine changes in the mean global physical parameters of MW Lyr during its Blazhko cycle in \cite{mw2}.

DM Cyg, a short period ($P=0.42$ d) RRab star was selected to be observed in the frame of the Konkoly Blazhko Survey because it had been announced to show phase modulation with a period of 26 d  by \cite{lf}, but neither the NSVS data \citep{nsvs} nor the complete data set of its maximum timings show phase modulation with this period  \citep{ibvs}.

Though DM Cyg is a relatively bright ($V=10-11$ mag) RR Lyrae star,  complete, accurate, multicolour light curve of its  pulsation has never been published. Its period change was, however, regularly monitored by different groups of observers. The GEOS database\footnote{\tt {http://dbrr.ast.obs-mip.fr/maxRR.html}} lists 260 maximum times of DM Cyg between 1900 and 2008. 

In the present paper we publish our extended CCD observations of DM Cyg and the results of the analysis of its light curve modulation. Archive photographic and photoelectric Konkoly data are also processed.

\section{Data}

\begin{table}
\caption{Konkoly photometric observations of DM Cyg. }
\label{photdata}
\begin{tabular}{lccc}
\hline
 HJD & mag$^{*}$ & detector & filter  \\
$2400000 +$& &&\\
\hline
\hline
54265.45479    & $-0.249$& CCD & V \\
54265.46045    & $-0.404$& CCD & V \\
54265.46611    & $-0.501$& CCD & V \\
...&... &...&...\\
\hline
\multicolumn{4}{l}{$^*$ Relative magnitudes for the CCD and photoelectric data. }\\
\hline

\end{tabular}
\end{table}
\begin{table}
\caption{Normal maximum timings derived from the Konkoly photographic, photoelectric and CCD observations.}
\label{maxtime}
\begin{tabular}{clc}
\hline
 JD & Normal maximum timings & obs. \\
\hline
$2427667-2429226$ & $2429078.482$ & pg \\
$2434186-2436373$ & $2435342.347$ & pg \\
$2443747-2444133$ & $2443777.333$ & pe \\
$2454265-2454407$ & $2454334.3467$ & CCD \\
$2454597-2454711$ & $2454661.4210$ & CCD \\
\hline
\end{tabular}
\end{table}

CCD observations were obtained with the automated 60 cm telescope of the Konkoly Observatory, Sv\'abhegy, Budapest equipped with a Wright Instruments $750\times1100$ CCD camera and $BVI_C$ filters. Measurements were taken on 81 nights between July 2007 and Sept 2008. About 3100 data points in each band were gathered. Exposition times were 200, 60 and 40 sec or a bit longer depending on the sky transparency in the $BVI_C$ bands, respectively. Data reduction was performed using standard IRAF\footnote{{\sc IRAF} is distributed by the National Optical Astronomy Observatories, which are operated by the Association of Universities for Research in Astronomy, Inc., under cooperative agreement with the National Science Foundation.} packages. Aperture photometry of DM Cyg (21:21:11.548 +32:11:28.71) and several neighbouring stars were carried out in order to check the stability of the photometry and the constancy of the comparison stars. The relative magnitudes of DM~Cyg  measured to the mean magnitudes of C1=GSC2.2  N0330220980 (21:21:30.731 +32:13.05.78), and C2=GSC2.2 N03302207371 (21:20:59.483 +32:13.01.63) are used in the analysis. The Tycho $B$ and $V$ magnitudes of the comparison stars C1 and C2 are $B_T=12.705, V_T=11.799$, and $B_T=13.081, V_T=12.150$, respectively \citep{tycho}. Second order extinction correction of the data were applied in the $B$ band. Relative magnitudes are transformed to standard $BVI_C$ magnitudes. More details on the reduction procedure are given in \cite{mw1}.

Archive photoelectric and photographic data obtained with the 60 cm telescope in 1978 and with a 16 inch astrograph between 1934 and 1958 are also utilized. Photoelectric observations were obtained on 4 nights, the $BV$ magnitudes were measured relative to GSC2.2 N03302207371.  The photographic measurements comprise data from 40 nights, the plates were evaluated using GSC2.2 $B$ magnitudes of the surrounding stars.

\begin{figure}
\centering
  \includegraphics[width=9cm]{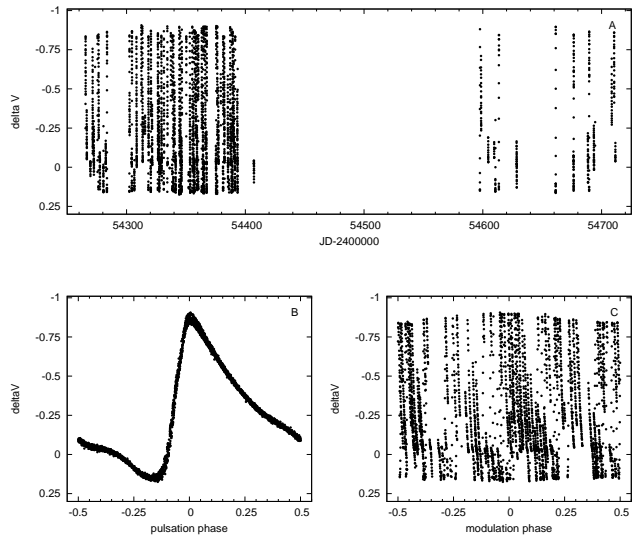}
  \caption{Delta $V$ magnitudes versus Julian Date, data phased with the  0.419863\,d pulsation and the \hbox{10.57\,d} modulation periods are shown in panels A, B and C, respectively. }
\label{lc}
\end{figure}

\begin{figure}
\centering
  \includegraphics[width=9.3cm]{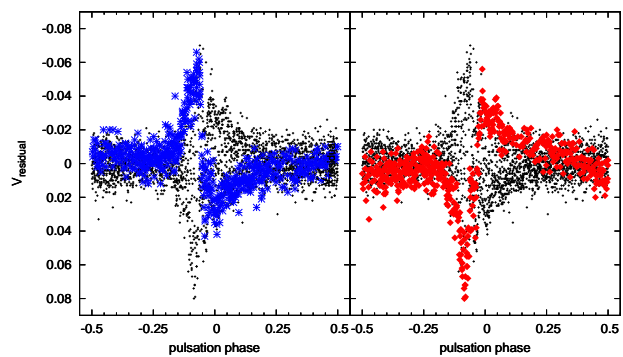}
  \caption{Residual $V$ light curve  of DM~Cyg after removing the pulsation components from the data according to the light curve solution given in Table~\ref{big}.  Residuals of the smallest (stars) and the largest (diamonds) amplitude phases are shown with large symbols in the left-hand and right-hand panels, respectively.  Note that the residuals have larger amplitude at around the middle of the rising branch of the pulsation ($phase=-0.1$) than around pulsation maxima ($phase=0$) due to phase modulation.}
\label{lepke}
\end{figure}

Table~\ref{photdata} shows a sample of the Konkoly photometric data of DM Cyg. All the photometric observations are available in the online version of the journal as Supplementary Material. 

Seasonal normal maximum timings from each data set have been determined for five epochs, these data are listed in Table~\ref{maxtime}.

\section{Results}

\begin{figure*}
\centering
\includegraphics[width=14.1cm]{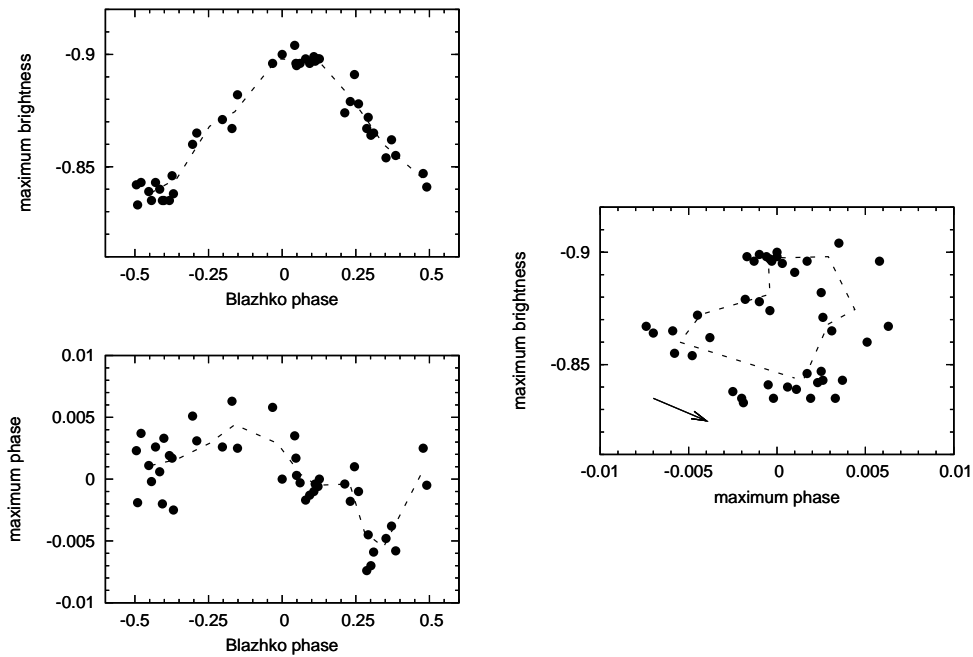}
\caption{Maximum $V$ brightness and maximum phase values phased according to the  Blazhko period are plotted in the top-left and bottom-left panels, respectively. Their typical errors are 0.005 mag and 0.002-0.004 phase. The right-hand panel shows the maximum brightness vs maximum phase data. The progression of data during the Blazhko cycle in this diagram is anti-clockwise. Dashed lines connect the mean values in 10 phase bins of the modulation.}
\label{max-oc}
\end{figure*}

The accurate CCD observations reveal that the light curve of DM Cyg is not stable. It shows small amplitude modulation, the variations in maximum brightness and phase are only about 0.07 mag and 0.005 d (7 min, 0.012 pulsation phase), respectively.

The $V$ light curve of DM Cyg and the results obtained folding the data with the pulsation and modulation periods are shown in the three panels of Fig.~\ref{lc}. The phased residual light curve after the removal of the mean pulsation light variation is given in Fig.~\ref{lepke}. This figure shows the differences of the amplitude of the modulation at different phases of the pulsation. The well observable amplitude of the modulation is concentrated to a narrow phase range of the pulsation around minimum, rising branch and maximum. A similar behaviour has been already found in RR Gem and SS Cnc, in two other small modulation amplitude Blazhko variables \citep{rrgI,ssc}. The amplitude of the the residual variation is the largest at around the middle of the rising branch of the pulsation light curve ($phase =-0.1$) indicating phase modulation of the rising branch as well. There is a difference between the phase of the amplitude modulation of the maximum of the light curve and the phase of the phase modulation of the same feature. In particular the maximum positive displacement of the timings of the maximum brightness precedes the occurrence of the brightest maximum by about 0.25 Blazhko phase as it is shown Fig.~\ref{max-oc}.

The elements of the pulsation and the modulation are: 

\begin{equation}
T_{\mathrm{max\,puls}} = 2\,454\,312.514\,{\mathrm [HJD]} + 0.419863\cdot E_{\mathrm{puls}},
\end{equation}

and

\begin{equation}
T_{\mathrm{max\,Bl}} = 2\,454\,312.514\,{\mathrm [HJD]} + 10.57 \cdot E_{\mathrm{Bl}}.
\end{equation}

\noindent
{$E_{\mathrm{puls}}$ and $E_{\mathrm{Bl}}$ denote the epoch number of the pulsation and modulation cycles, respectively.}

The pulsation and modulation periods correspond to the mean values of the frequencies of the  light curve solutions of the $B, V$ and $I_C$ data  using `locked' frequency solutions of the $kf_0$ pulsation and $kf_0\pm f_\mathrm{m}$,  and $f_\mathrm{m}$ modulation  frequency components  (i.e., the modulation side lobe components are at the positions of the linear combination frequencies). 

Data analysis was performed using the different applications of the MUFRAN package \citep{mufran}, a linear combination fitting program developed by \'A. S\'odor, and the linear and nonlinear curve fitting abilities of gnuplot\footnote{\tt http://www.gnuplot.info/}.

\begin{figure*}
  \centering
  \includegraphics[width=8cm,angle=-90]{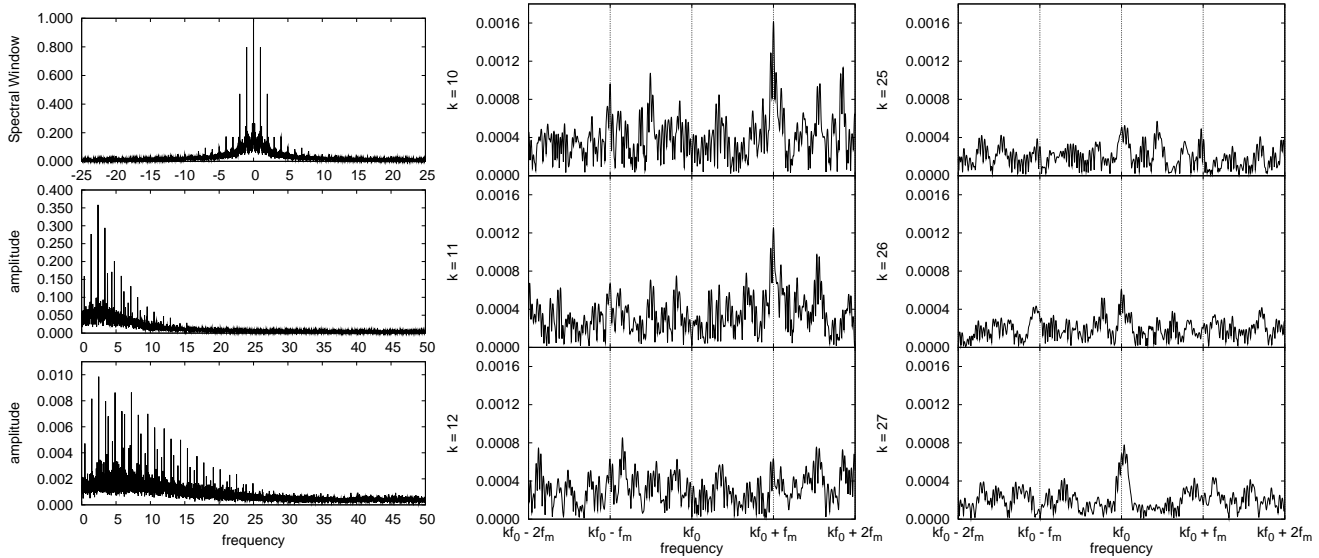}
  \caption{The left-hand panels show the spectral window, the amplitude spectrum  and the residual spectrum after the removal of the pulsation components for the CCD  $V$ data. The middle and right-hand panels show residual spectra  in the vicinity of  the $10f_0, 11f_0,$ and $12f_0$ ($k=10,11,12$) and of the  $25f_0, 26f_0,$ and $27f_0$ ($k=25,26,27$) frequencies, respectively. The spectrum has been prewhitened for the pulsation components up to the 24th order, and in the right hand panels for the detected modulation components, too. These spectra indicate that modulation frequency components are present only in the first 10-11 orders, while the pulsation can be accurately described if harmonic frequency components are taken into account up to as high as the 27th order. For clarity, vertical grids denote the positions of the $kf_0, kf_0+f_m$ and $kf_0-f_m$ frequencies.}
  \label{spektrum}
\end{figure*}

\begin{table*}
\caption{Fourier amplitudes and phases of the pulsation and modulation frequencies in DM Cyg. Initial epoch corresponds to maximum of both the pulsation and the modulation light variations as given in Eqs. 1 and 2.}
\label{big}
\begin{tabular}{lcccccccccc}
\hline
 & frequency & \multicolumn{2}{c}{$B$} & \multicolumn{2}{c}{$V$} & \multicolumn{2}{c}{$I_c$} & \multicolumn{2}{c}{$V_{tt}^{*}$}&$\sigma_{\varphi(V_{tt})}$\\
 & & amp & $\varphi$ & amp & $\varphi$ & amp & $\varphi$ & amp & $\varphi$ &\\
\hline
$f_m$ & $0.094607$ & $0.0026$ & $1.678$ & $0.0017$ & $1.637$ & $0.0014$ & $1.517$ & $0.0017$ & $1.644$ & $0.107$ \\ 
$f_0 - 2f_m$ & $2.192515$ & $0.0015$ & $3.363$ & $0.0008$ & $2.989$ & $0.0005$ & $3.939$ & $0.0008$ & $2.697$ & $0.211$ \\ 
$f_0 - f_m$ & $2.287122$ & $0.0036$ & $2.694$ & $0.0022$ & $2.618$ & $0.0011$ & $2.839$ & $0.0048$ & $3.675$ & $0.037$ \\ 
$f_0$ & $2.381729$ & $0.4903$ & $3.944$ & $0.3532$ & $3.874$ & $0.2170$ & $3.670$ & $0.3532$ & $3.874$ & $0.001$ \\ 
$f_0 + f_m$ & $2.476337$ & $0.0132$ & $3.822$ & $0.0096$ & $3.854$ & $0.0061$ & $3.975$ & $0.0057$ & $4.043$ & $0.032$ \\ 
$2f_0 - 2f_m$ & $4.574244$ & $0.0014$ & $0.664$ & $0.0008$ & $0.275$ & $0.0004$ & $0.976$ & $0.0008$ & $0.573$ & $0.219$ \\ 
$2f_0 - f_m$ & $4.668851$ & $0.0035$ & $2.355$ & $0.0026$ & $2.404$ & $0.0012$ & $2.513$ & $0.0043$ & $3.757$ & $0.042$ \\ 
$2f_0$ & $4.763459$ & $0.2598$ & $4.109$ & $0.1919$ & $4.090$ & $0.1189$ & $4.029$ & $0.1920$ & $4.090$ & $0.001$ \\ 
$2f_0 + f_m$ & $4.858066$ & $0.0105$ & $3.511$ & $0.0080$ & $3.572$ & $0.0049$ & $3.619$ & $0.0038$ & $3.284$ & $0.047$ \\ 
$3f_0 - f_m$ & $7.050581$ & $0.0050$ & $3.173$ & $0.0034$ & $3.187$ & $0.0024$ & $3.270$ & $0.0053$ & $4.058$ & $0.034$ \\ 
$3f_0$ & $7.145188$ & $0.1545$ & $4.475$ & $0.1160$ & $4.476$ & $0.0728$ & $4.466$ & $0.1162$ & $4.476$ & $0.002$ \\ 
$3f_0 + f_m$ & $7.239795$ & $0.0100$ & $3.928$ & $0.0079$ & $3.988$ & $0.0048$ & $4.003$ & $0.0040$ & $3.772$ & $0.044$ \\ 
$4f_0 - f_m$ & $9.432310$ & $0.0024$ & $3.179$ & $0.0019$ & $3.237$ & $0.0012$ & $3.230$ & $0.0027$ & $4.562$ & $0.065$ \\ 
$4f_0$ & $9.526917$ & $0.0816$ & $4.970$ & $0.0628$ & $4.971$ & $0.0396$ & $4.972$ & $0.0629$ & $4.970$ & $0.003$ \\ 
$4f_0 + f_m$ & $9.621525$ & $0.0086$ & $4.464$ & $0.0063$ & $4.437$ & $0.0042$ & $4.466$ & $0.0036$ & $4.222$ & $0.050$ \\ 
$5f_0 - f_m$ & $11.814039$ & $0.0019$ & $3.800$ & $0.0016$ & $3.799$ & $0.0010$ & $3.646$ & $0.0027$ & $4.826$ & $0.067$ \\ 
$5f_0$ & $11.908647$ & $0.0510$ & $5.142$ & $0.0391$ & $5.171$ & $0.0247$ & $5.204$ & $0.0392$ & $5.173$ & $0.004$ \\ 
$5f_0 + f_m$ & $12.003254$ & $0.0069$ & $4.868$ & $0.0053$ & $4.889$ & $0.0035$ & $4.862$ & $0.0030$ & $4.874$ & $0.060$ \\ 
$6f_0 - f_m$ & $14.195768$ & $0.0012$ & $4.312$ & $0.0011$ & $4.130$ & $0.0008$ & $4.502$ & $0.0019$ & $5.325$ & $0.094$ \\ 
$6f_0$ & $14.290376$ & $0.0338$ & $5.592$ & $0.0255$ & $5.617$ & $0.0162$ & $5.665$ & $0.0257$ & $5.617$ & $0.007$ \\ 
$6f_0 + f_m$ & $14.384983$ & $0.0056$ & $5.159$ & $0.0045$ & $5.169$ & $0.0027$ & $5.102$ & $0.0028$ & $5.047$ & $0.064$ \\ 
$7f_0 - f_m$ & $16.577498$ & $0.0015$ & $5.147$ & $0.0010$ & $4.585$ & $0.0010$ & $4.748$ & $0.0015$ & $5.503$ & $0.117$ \\ 
$7f_0$ & $16.672105$ & $0.0185$ & $5.861$ & $0.0146$ & $5.921$ & $0.0088$ & $5.979$ & $0.0147$ & $5.925$ & $0.012$ \\ 
$7f_0 + f_m$ & $16.766713$ & $0.0040$ & $5.601$ & $0.0034$ & $5.591$ & $0.0020$ & $5.667$ & $0.0022$ & $5.527$ & $0.080$ \\ 
$8f_0 - f_m$ & $18.959227$ & $0.0003$ & $0.174$ & $0.0006$ & $5.416$ & $0.0004$ & $6.238$ & $0.0012$ & $6.113$ & $0.143$ \\ 
$8f_0$ & $19.053834$ & $0.0114$ & $6.253$ & $0.0089$ & $0.030$ & $0.0050$ & $0.167$ & $0.0089$ & $0.033$ & $0.018$ \\ 
$8f_0 + f_m$ & $19.148442$ & $0.0030$ & $5.913$ & $0.0026$ & $5.954$ & $0.0016$ & $6.054$ & $0.0018$ & $5.879$ & $0.099$ \\ 
$9f_0 - f_m$ & $21.340956$ & $0.0009$ & $0.444$ & $0.0006$ & $0.525$ & $0.0009$ & $0.135$ & $0.0010$ & $0.647$ & $0.172$ \\ 
$9f_0$ & $21.435564$ & $0.0048$ & $0.309$ & $0.0037$ & $0.482$ & $0.0024$ & $0.769$ & $0.0038$ & $0.480$ & $0.046$ \\ 
$9f_0 + f_m$ & $21.530171$ & $0.0021$ & $0.204$ & $0.0019$ & $0.034$ & $0.0013$ & $0.248$ & $0.0015$ & $6.239$ & $0.113$ \\ 
$10f_0 - f_m$ & $23.722686$ & $0.0004$ & $1.793$ & $0.0005$ & $1.018$ & $0.0008$ & $0.687$ & $0.0006$ & $1.077$ & $0.289$ \\ 
$10f_0$ & $23.817293$ & $0.0016$ & $0.291$ & $0.0012$ & $0.760$ & $0.0007$ & $2.221$ & $0.0012$ & $0.780$ & $0.141$ \\ 
$10f_0 + f_m$ & $23.911900$ & $0.0015$ & $0.576$ & $0.0013$ & $0.624$ & $0.0009$ & $0.652$ & $0.0011$ & $0.573$ & $0.162$ \\ 
$11f_0$ & $26.199022$ & $0.0006$ & $4.556$ & $0.0009$ & $4.183$ & $0.0009$ & $3.499$ & $0.0009$ & $4.147$ & $0.192$ \\ 
$11f_0 + f_m$ & $26.293630$ & $0.0009$ & $1.438$ & $0.0010$ & $0.835$ & $0.0008$ & $1.374$ & $0.0011$ & $0.818$ & $0.162$ \\ 
$12f_0$ & $28.580752$ & $0.0019$ & $4.559$ & $0.0014$ & $4.526$ & $0.0015$ & $4.265$ & $0.0015$ & $4.462$ & $0.116$ \\ 
$13f_0$ & $30.962481$ & $0.0024$ & $4.971$ & $0.0021$ & $4.891$ & $0.0017$ & $4.564$ & $0.0021$ & $4.878$ & $0.082$ \\ 
$14f_0$ & $33.344210$ & $0.0026$ & $5.317$ & $0.0024$ & $5.239$ & $0.0019$ & $5.240$ & $0.0024$ & $5.185$ & $0.073$ \\ 
$15f_0$ & $35.725940$ & $0.0023$ & $5.712$ & $0.0023$ & $5.560$ & $0.0016$ & $5.429$ & $0.0024$ & $5.521$ & $0.073$ \\ 
$16f_0$ & $38.107669$ & $0.0024$ & $6.073$ & $0.0022$ & $5.883$ & $0.0020$ & $5.677$ & $0.0022$ & $5.845$ & $0.079$ \\ 
$17f_0$ & $40.489398$ & $0.0021$ & $6.242$ & $0.0019$ & $6.154$ & $0.0015$ & $6.173$ & $0.0019$ & $6.104$ & $0.088$ \\ 
$18f_0$ & $42.871127$ & $0.0019$ & $0.187$ & $0.0017$ & $0.166$ & $0.0015$ & $0.062$ & $0.0017$ & $0.106$ & $0.096$ \\ 
$19f_0$ & $45.252857$ & $0.0016$ & $0.452$ & $0.0014$ & $0.546$ & $0.0015$ & $0.444$ & $0.0015$ & $0.533$ & $0.111$ \\ 
$20f_0$ & $47.634586$ & $0.0018$ & $0.765$ & $0.0014$ & $0.920$ & $0.0011$ & $0.755$ & $0.0014$ & $0.881$ & $0.121$ \\ 
$21f_0$ & $50.016315$ & $0.0009$ & $0.818$ & $0.0012$ & $1.023$ & $0.0011$ & $1.094$ & $0.0013$ & $0.947$ & $0.132$ \\ 
$22f_0$ & $52.398045$ & $0.0014$ & $1.161$ & $0.0010$ & $1.394$ & $0.0010$ & $1.384$ & $0.0011$ & $1.321$ & $0.162$ \\ 
$23f_0$ & $54.779774$ & $0.0009$ & $1.432$ & $0.0010$ & $1.644$ & $0.0007$ & $1.658$ & $0.0010$ & $1.673$ & $0.177$ \\ 
$24f_0$ & $57.161503$ & $0.0009$ & $2.064$ & $0.0010$ & $2.306$ & $0.0006$ & $2.073$ & $0.0010$ & $2.259$ & $0.177$ \\ 
$25f_0$ & $59.543233$ & $0.0006$ & $2.156$ & $0.0008$ & $2.556$ & $0.0007$ & $2.521$ & $0.0008$ & $2.568$ & $0.226$ \\ 
$26f_0$ & $61.924962$ & $0.0010$ & $2.759$ & $0.0005$ & $2.907$ & $0.0004$ & $2.592$ & $0.0004$ & $2.883$ & $0.402$ \\ 
$27f_0$ & $64.306691$ & $0.0007$ & $2.871$ & $0.0006$ & $2.946$ & $0.0007$ & $3.139$ & $0.0006$ & $2.961$ & $0.258$ \\ 
\hline
\multicolumn{11}{l}{$^*$ Fourier parameters of the time transformed $V$ data}\\
\hline
\end{tabular}
\end{table*}

\subsection{The light curve solution}

\begin{figure}
  \centering
  \includegraphics[width=7.8cm,]{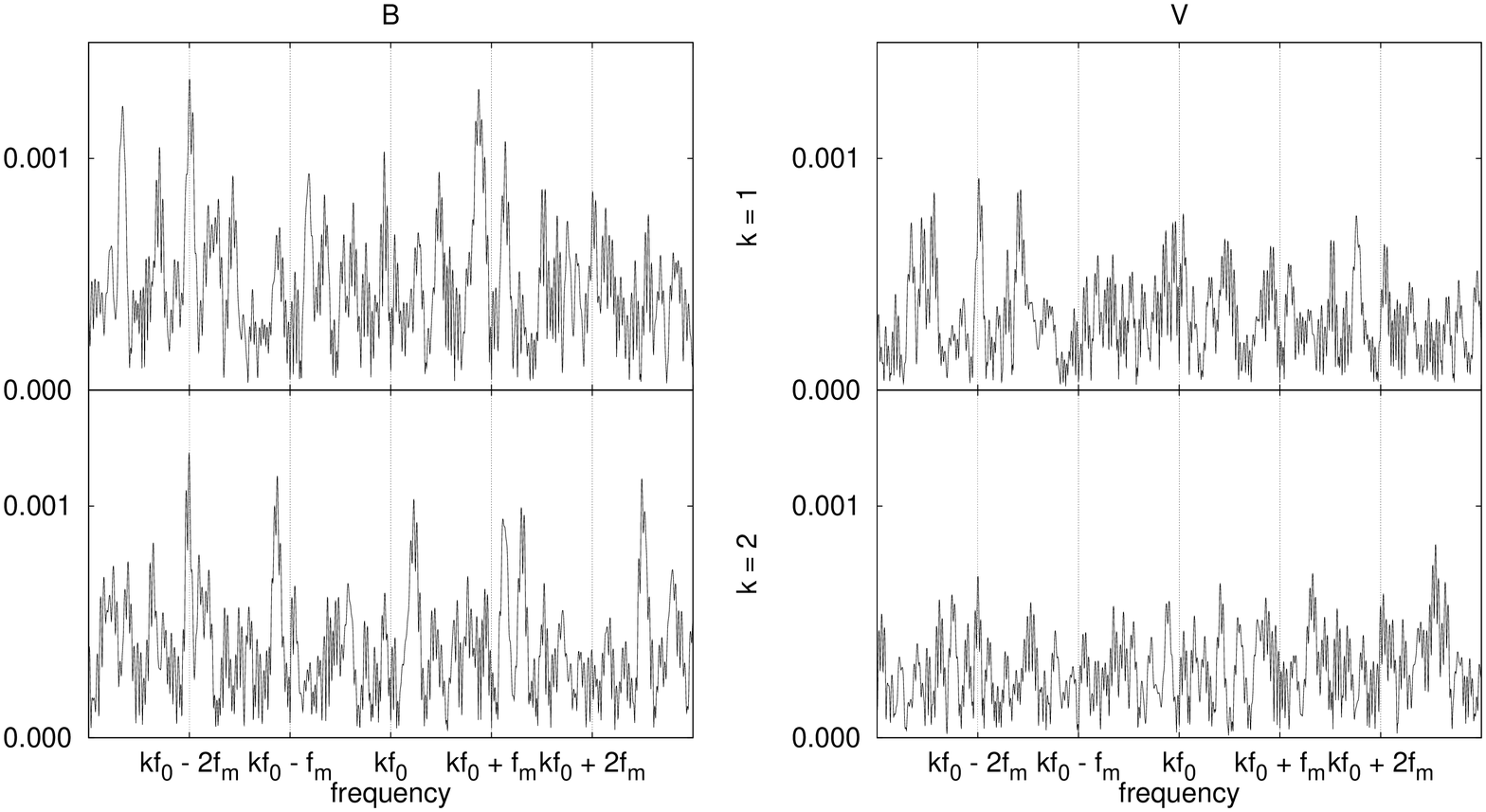}
  \caption{Fourier spectra of the $V$ and $B$ residual data in the vicinity of $f_0$ and $2f_0$. The pulsation and the $kf_0\pm f_m$ modulation frequencies have been removed. The $f_0 - 2f_m$ and $2f_0 - 2f_m$ quintuplet components appear in these residual spectra with \mbox{$\sim0.001$ mag} amplitude.}
  \label{kvint}
\end{figure}

\begin{figure}
\centering
\includegraphics[width=8cm]{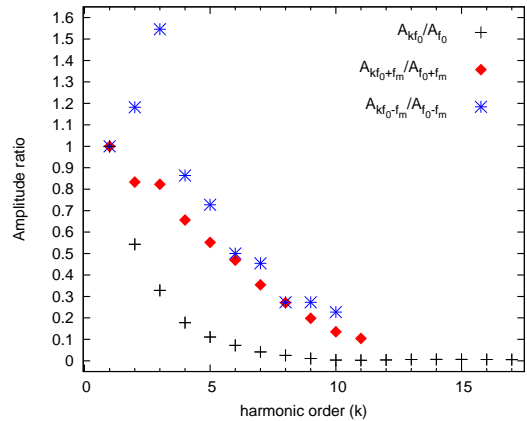}
\caption{Amplitude decrease of the pulsation and modulation side lobe frequencies at different orders shown as amplitudes normalised to the amplitudes of the first order (k=1) components. The decrease of the amplitudes of the harmonic components of the pulsation is exponential like, while the decrease of the amplitudes of the side lobe frequencies is more linear, with irregular character in the low orders.}
\label{amp}
\end{figure}

The Fourier amplitudes and phases of the pulsation and modulation frequency components identified in the spectra of the $BVI_C$ light curves of DM Cyg are summarized in Table~\ref{big}. The Fourier decompositions use $sin$ terms and the initial epoch corresponds to one of the brightest maxima of the modulated light curve, $T0=2454312.514$.  The errors of the amplitudes are $\sim0.0002$ mag, the errors of the phases for the time transformed $V$ data (see the details later) are given in the last column. The pulsation components are detectable up to the 27th order, while the modulation side frequencies ($kf_0\pm f_m$) diminish at around the 11th order as documented in Fig.~\ref{spektrum}. The modulation frequency $f_m$ is present without question in  each of the  $B, V, I$ spectra with similar amplitudes as the 4th order negative and the 9th order positive modulation side lobe components have. Quintuplet frequencies are detected at $f_0- 2f_m$ and $2f_0 - 2f_m$ as shown in Fig.~\ref{kvint}.

The 51 frequencies listed in Table~\ref{big} fit the data with 6-8 mmag residual scatter in each band, which is about the level of the observational noise. No further periodic or stochastic variation in the residuals is detected. 
 
The decrease of the amplitudes of the pulsation components  with increasing order is smooth and exponential-like. The amplitude decrease of the modulation components is however, different. The amplitudes of the low order modulation components behave irregularly, while they show a linear decrease at higher orders (see Fig.~\ref{amp}). Similar behaviour of the amplitude decrease of the detected frequencies were demonstrated for RR Gem, SS Cnc and SS For \citep{rrgI,ssc,ssfor}.

\begin{figure*}
  \centering
  \includegraphics[width=8.5cm,angle=-90]{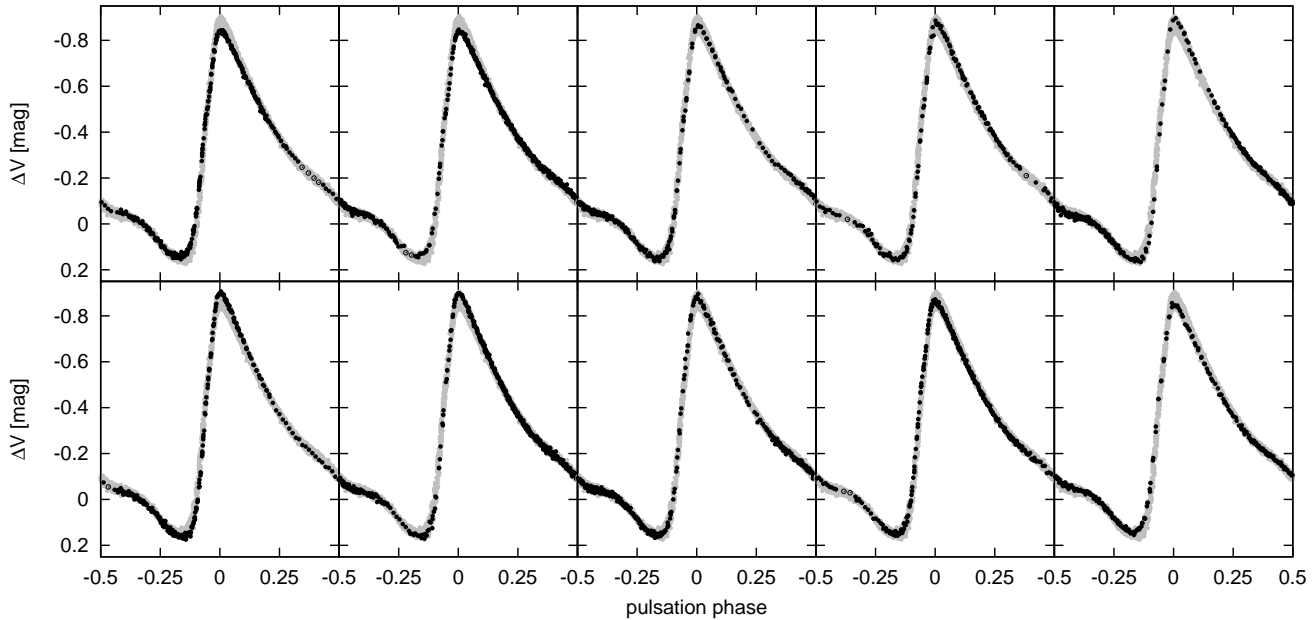}
  \caption{$\Delta V$ light curves in 10 phase bins of the modulation. All the observations are also shown with gray colour in each plot. Empty circles denote artificial data which are used to stabilize the Fourier fits if there is some gap in the observations.  }
  \label{lck}
\end{figure*}

\begin{figure}
  \centering
  \includegraphics[width=8.5cm]{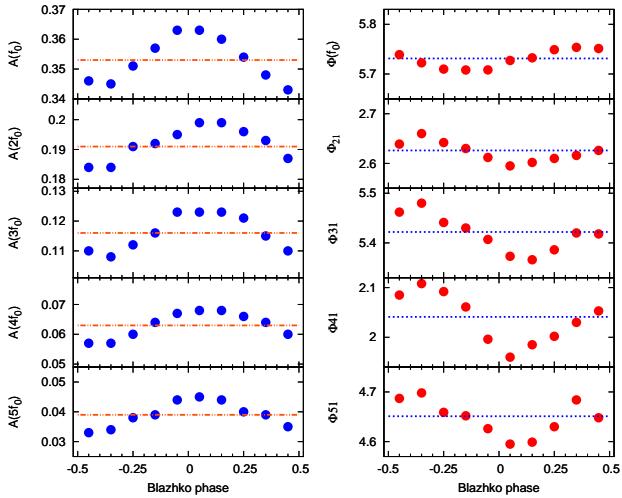}
  \caption{Fourier amplitudes of the first 5 harmonic components of the $V$ light curves  (left-hand panels) and the phase of the $f_0$ pulsation frequency and the epoch independent phase differences (right-hand panels) at 10 different phases of the Blazhko cycle. Note, that the scales of each amplitude and phase plots are identical.}
  \label{param}
\end{figure}

Fig.~\ref{lck} shows the $V$ light curves belonging to 10 different phase bins of the modulation. The first five Fourier parameters of the 15th order harmonic fits to these individual $V$ light curves are plotted in Fig.~\ref{param}. Comparing the variations of the Fourier parameters of the light curves of DM Cyg with that of MW Lyr \citep[Fig. 12 in][]{mw1} the following differences are conspicuous. The amplitudes of the detected changes in the Fourier parameters are about one order of magnitude smaller than the amplitudes of the light curve variations in MW Lyr. The amplitudes of the variation of the amplitudes of the $f_0,...5f_0$ components decrease more drastically with increasing order in MW Lyr than in DM Cyg. This is most probably connected to that the pulsation components are detected only up to the 12th order  in MW Lyr, while they can be observed up to the 27th order in DM Cyg. Nevertheless the phase relations between the amplitude modulation and the variation in the phase of the $f_0$ pulsation component are the same for the two stars, the detected changes in the epoch independent phase differences [$\varphi(f_{k1})$] are, however, significantly different. While in DM Cyg the $\varphi(f_{k1})$ phase differences show sinusoidal variations with 0.25 ($\pi/2$) phase shift and with larger amplitude than $\varphi(f_0)$, in MW Lyr the amplitudes of the variations of the $\varphi(f_{k1})$ components are smaller than the amplitude of $\varphi(f_0)$ and they do not show strictly regular behaviour with the modulation period.

\begin{figure}
  \centering
  \includegraphics[width=4.4cm,angle=-90]{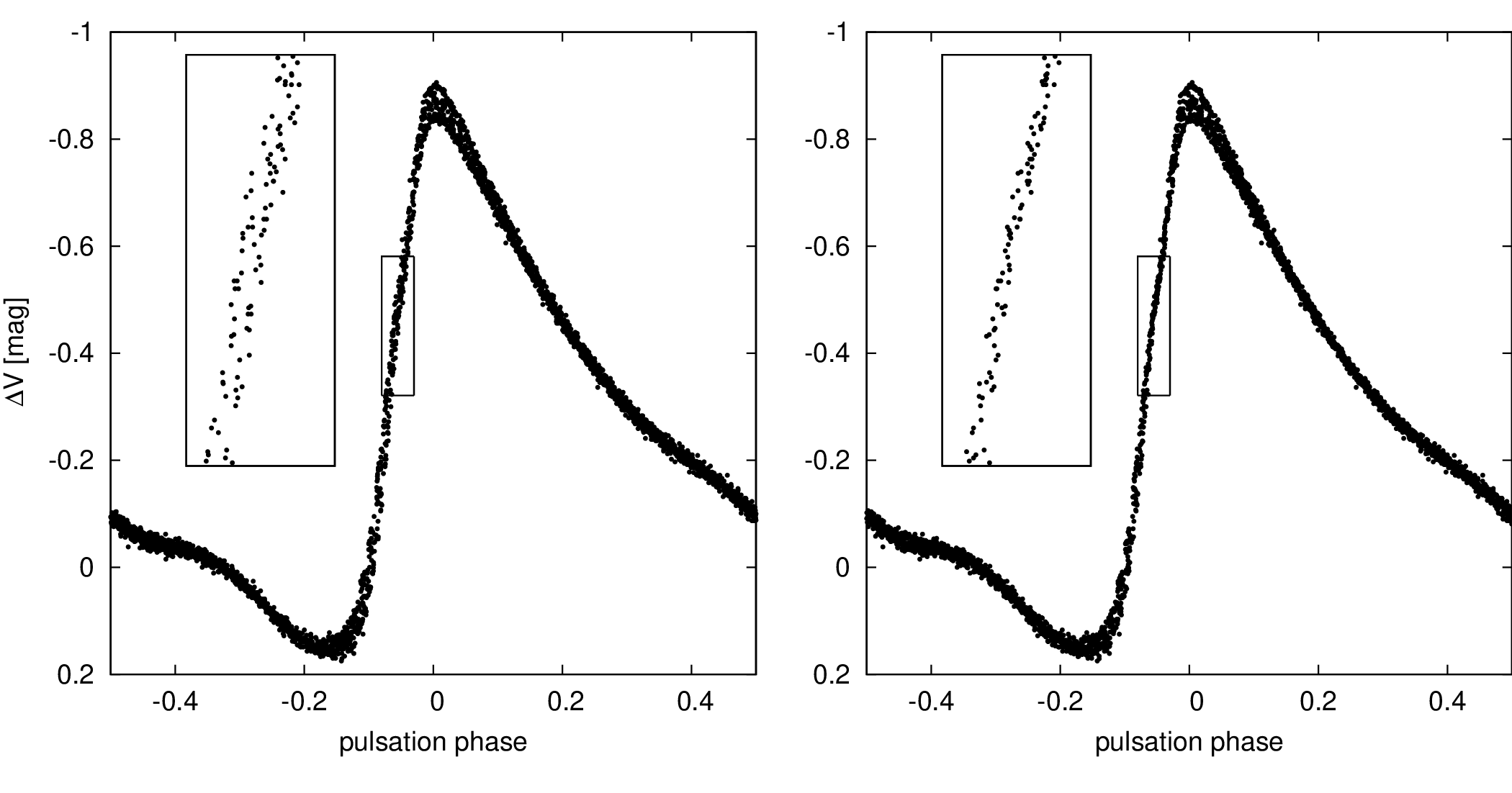}
  \caption{Folded light curve of the $V$ observations (left-hand panel) and that of the time transformed data (right-hand panel). The times of the observations are corrected according to the phase variation of the $f_0$ pulsation component during the modulation cycle in the transformed data. The inserts magnify the plots at around the middle of the rising branch. The original data show significant spread of the observations here due to phase modulation, while in the time transformed data the phase modulation is eliminated as the  narrowness of the rising branch indicates.  }
  \label{tt}
\end{figure}

In \cite{mw1} we have introduced a new method of analysing Blazhko variables' light curves. It was shown that if the times of the observations are corrected according to an appropriate time transformation defined by the variation of the phase of $f_0$ during the Blazhko cycle, then the modulation can be separated into phase and amplitude modulation components. The time transformed data showed pure amplitude modulation for MW Lyr. Though the modulation amplitude of DM Cyg is very small, we have checked how a similar time transformation influences its modulation properties. Transforming the times of the observations according to a continuous harmonic function of the $\varphi(f_0)$  data, the $V$ observations folded with the pulsation period is compared to the folded light curve of the original data of DM Cyg in Fig.~\ref{tt}. Enlarged plots of the middle of the rising branch are inserted in the figures. Though the difference between the original and the time transformed data is small, the reduction of the phase modulation component in the time transformed data is indicated  by the narrowness of the rising branch in the right-hand panel of Fig.~\ref{tt}.

\begin{figure}
  \centering
  \includegraphics[width=8.7cm]{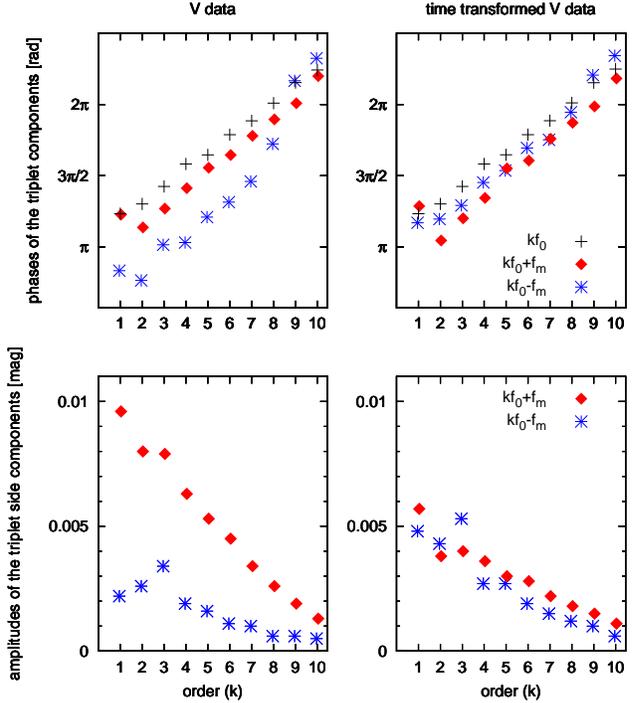}
  \caption{Phases of the pulsation ($kf_0$) and modulation side lobe  ($kf_0\pm f_m$) frequencies for the original and the time transformed $V$ data sets are shown in the top panels. The phase differences between the three components of the triplets remain close the same in the different orders with larger deviations from these constant values in the $k=1,2$ and in the highest orders in the original $V$ data (left-hand panel). The phases of the components of the triplets have nearly the same values in the different orders  in the time transformed data (right-hand panel). Though the phases of the different frequency components are epoch dependent, the rough constancy of the phase differences between the triplet components at different orders holds for any epoch both for the original and the time transformed data. According to model calculations, the phase differences are close to zero (as observed in the case of the time transformed data) if only amplitude modulation occurs and the initial phase corresponds to the maximum amplitude of the modulation. Bottom panels show the amplitudes of the side lobe frequencies for the original and the time transformed data. The $kf_0 + f_m$ components have systematically larger amplitudes than the $kf_0 - f_m$ components have in the original data as shown in the bottom left panel. Contrarily, the amplitudes of the side frequencies are very similar in each order in the time transformed data (bottom right panel), as expected if the modulation is pure amplitude modulation.}
  \label{phase}
\end{figure}

In the last three columns of Table~\ref{big} the Fourier amplitudes and phases and the errors of the phases are listed for the time transformed $V$ data.  The amplitudes of the modulation side frequency components $kf_0 + f_m$ and $kf_0 - f_m$ are about the same and the phases of the triplets in the different orders show definite phase coherency according to the Fourier solution of the time transformed data (see Fig.~\ref{phase}). 
As the phases of the components of the triplets are epoch dependent, this phase coherency holds only if the initial epoch corresponds to the phase of the maximum of the modulation. It can be proved analytically that the phases of the components of a modulation frequency-triplet are identical if the modulation is pure amplitude modulation and the initial epoch corresponds to $\pi/2$ using sine terms, i.e, it is chosen to be at the maximum phase of the modulation \citep{szeidl}. The phase coherency of the triplets and the symmetry of the amplitudes of the side frequencies confirm that the modulation of the time transformed data is basically amplitude modulation.
 
Based on analytical and test results it was recently shown by \cite{szeidl} that the phase difference between the phases of the amplitude and phase modulations is connected to the difference between the squares of the amplitudes of the $f_0 + f_m$ and $f_0- f_m$ components. If the $f_0 + f_m$ component has larger amplitude than $f_0- f_m$ has, then the occurrence of the largest amplitude of the pulsation (maximum of the brightness maxima) precedes the occurrence of the largest delay of the maximum timings (largest positive $O-C$ value) and the direction of the progression in the maximum brightness--maximum light phase plot (right-hand  panel in Fig.~\ref{max-oc}) is anti-clockwise.  Contrarily, if the larger amplitude modulation components are the $kf_0- f_m$ frequencies, then the phase of the largest delay of the light curve  precedes the phase of the maximum amplitude. DM Cyg is an example for the former case, the amplitudes of the $kf_0 + f_m$ components are larger than the amplitudes of the  $kf_0 -f_m$ components, the maximum of the amplitude modulation precedes the phase of the largest delay of the maxima and the direction of the progression in the right-hand panel in Fig.~\ref{max-oc} is anti-clockwise.

\begin{figure*}
\centering
\includegraphics[width=14.1cm]{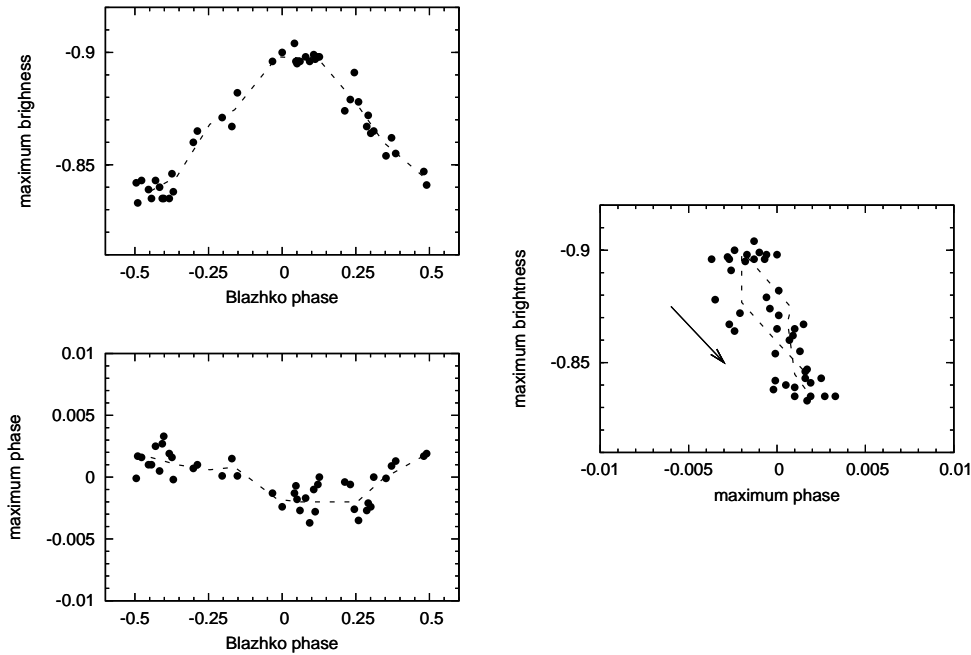}
\caption{The same figure as Fig.~\ref{max-oc} but for the time transformed data. The phase difference between the maximum brightness and maximum phase timings for the time transformed data is around $180 \degr$. The maximum brightness - maximum phase plot has been substantially compressed,  the direction of the progression is denoted by an arrow. Dashed lines connect the mean values in 10 phase bins of the modulation. }
  \label{max-oc.tt}
\end{figure*}

Due to the non-sinusoidal shape of the light curve and the residual phase modulation in the higher order pulsation components the phases of maxima  vary slightly in the time transformed data as well. The difference between the phase of the highest light maximum and the phase of the maximum brightness of the largest delay during the Blazhko cycle is around $180 \degr$ for this data set as it is shown in Fig.~\ref{max-oc.tt}, i.e. the largest delay of the light curve occurres when the amplitude of the pulsation is the smallest. Fig.~\ref{max-oc.tt} shows the same plots as Fig.~\ref{max-oc} but for the time transformed data. The right-hand panel shows already hardly any loop structure wider than the scatter of the data.

\subsection{Changes in the global physical parameters during the Blazhko cycle} 

In \cite{ipm} we demonstrated that from good quality multicolour photometric data the physical parameters of RRab stars can be determined with similar accuracy as with direct Baade-Wesselink analysis, using an inverse photometric method (IPM). We applied this method to the $BVI_C$ light curves of MW Lyr at different phases of its Blazhko modulation succesfully, about $1-2 \%$ changes in the mean global physical parameters ($L, T_{eff}, R$) were detected during the Blazhko cycle \citep{mw2}. 

Before analysing the light curves of DM Cyg at different phases of its modulation, first we have to determine those parameters that definitely do not vary during the Blazhko cycle. These are  the metallicity, the mass and the distance of the star and the dereddened standard magnitudes of the comparison stars i.e., the zero points of the magnitude scales used. 

The [Fe/H] of DM Cyg was determined spectroscopically by \cite{suntzeff} and \cite{layden}. The Fourier parameters of the mean $V$ light curve give [Fe/H]$=-0.01$  using the \citet[][Eq. 3]{jk} formula. As the spectroscopic observations correspond to $-0.16$ and 0.07 [Fe/H] values on the metallicity scale of the \cite{jk} formula, atmosphere models with [Fe/H]$=0.0$ and  $-0.1$ compositions \citep{kurucz} are used in the light curve modelling of DM Cyg.

The IPM finds mass values that are too large on evolutionary grounds in some cases if the mass is also allowed to vary by the fitting process. The fitting accuracy of the different mass solutions differs, however, only marginally.  The mean $B, V, I_C$ light curves of DM~Cyg are fitted with 5.1, 4.7 and 4.0 mmag residual scatters when the mass value is also allowed to vary. In this case the best solution is found at $\mathfrak{M} =0.84 \mathfrak{M}_{Sun}$. The r.m.s. of the fixed $\mathfrak{M} =0.55 \mathfrak{M}_{Sun}$ mass solution $B, V, I_C$ light curve fits are 5.2, 4.6 and 4.3 mmag, respectively. The average fitting accuracy in the three bands decreases only 0.1 mmag, from 4.7 to 4.6 mmag, when the mass is also allowed to be fitted. That is, the method is rather insensitive to the value of the mass. Therefore, in order to obtain reliable solutions, the IPM has been run with 0.50, 0.55 and 0.60 $\mathfrak{M}_{Sun}$ fixed mass values for the mean light curves of DM~Cyg.

\begin{table*}
\caption{Mean physical parameters of DM Cyg derived from its mean light curves using the IP method.}
\label{pp}
\begin{tabular}{lllllllll}
\hline
[Fe/H]&$\mathfrak{M/M}_{Sun}$&$ L/L_{Sun}$& $T_{eff}$& $R/R_{Sun}$&$M_V$ &$(B-V)_0$ & $(V-I)_0$ & d [pc] \\
 fixed&fixed&&&\\
\hline
 \,$\,\,0.0$&0.50& $35.7\pm1.5$ & $6510\pm55$& $4.66\pm0.09$&$0.97\pm0.05$&$0.43\pm0.02$&$0.49\pm0.01$&$1179\pm23$\\
&0.55& $38.0\pm1.8$ & $6540\pm40$& $4.76\pm0.09$&$0.90\pm0.04$&$0.44\pm0.01$&$0.48\pm0.01$&$1215\pm21$\\
&0.60& $39.0\pm1.4$ & $6535\pm40$& $4.85\pm0.08$&$0.86\pm0.04$&$0.42\pm0.01$&$0.48\pm0.01$&$1233\pm21$\\
\hline
$-0.1$&0.55& $38.0\pm1.0$ & $6533\pm31$& $4.78\pm0.09$&$0.90\pm0.03$&$0.45\pm0.01$&$0.48\pm0.01$&$1210\pm15$\\
\hline

\end{tabular}
\end{table*}
 \begin{figure*}
  \centering
  \includegraphics[width=17.5cm]{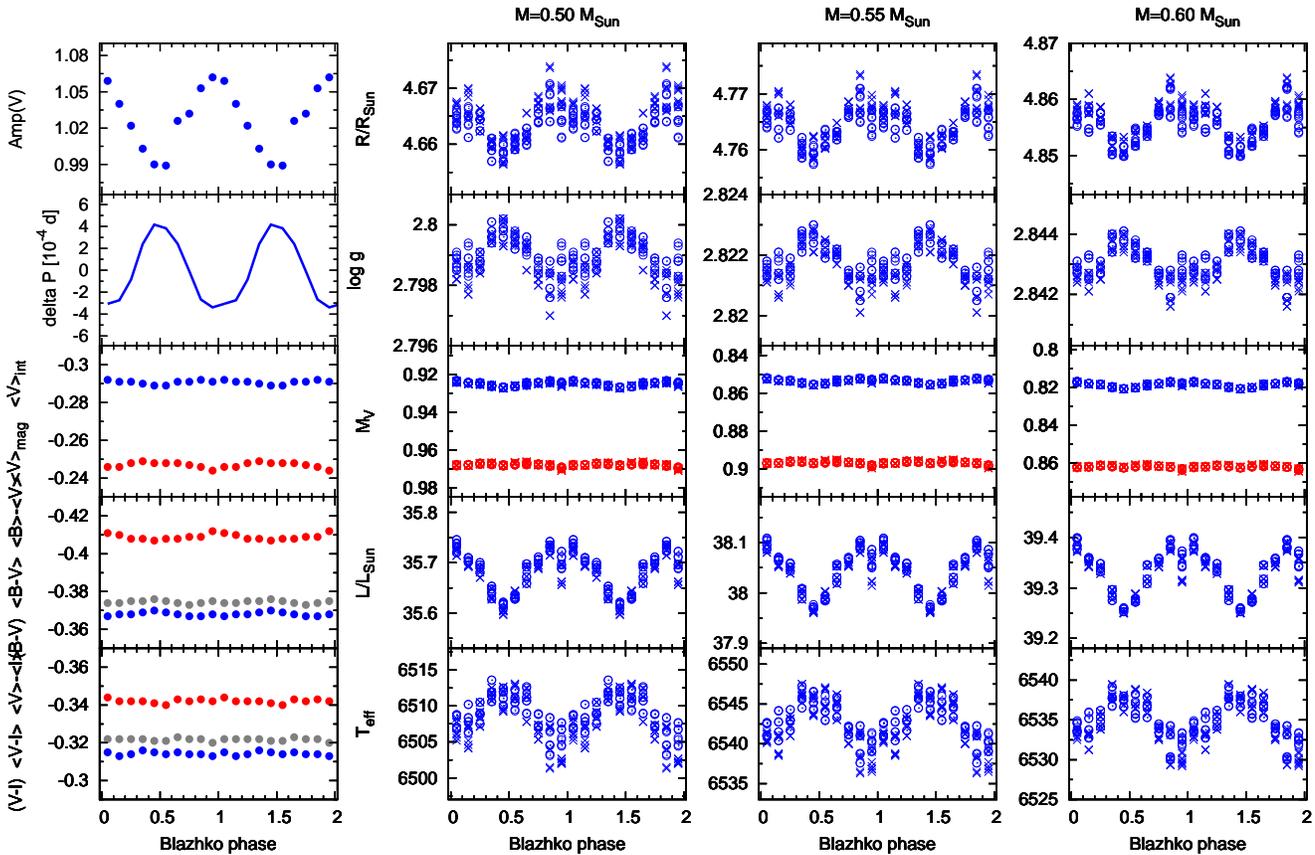}
  \caption{Variations of the observed mean (left-hand panels) and derived parameters of DM Cyg during the Blazhko cycle. Circles and crosses denote the results when the IP method use Liu's template radial velocity curve, and when the $V_{rad}$ curve is calculated from the $I_C$ light curve, respectively. No significant, systematic difference between the results can be seen depending on the choice of the initial radial velocity templates. The results of the IP method for fixed 0.50, 0.55 and 0.60 $\mathfrak{M/M}_{Sun}$ mass values show the same variations in each parameters. Only the absolute mean values of the parameters are different, according to the data given in Table~\ref{pp}. These results prove that the detected changes of the physical parameters during the Blazhko cycle of DM Cyg are not sensitive to the settings of the IP method, and are independent of the uncertainties of the values of the mean global parameters. See further details in the text.}
  \label{minden}
\end{figure*}

The global mean absolute physical parameters of DM Cyg averaged over both the pulsation and the Blazhko cycles and its distance are derived from the mean light curves using the IP method. The results are summarized in Table~\ref{pp} assuming different possible mass and [Fe/H] values. Note that the zero points of the colour and magnitude scales defined by the dereddened magnitudes of the comparison stars, are also allowed to vary i.e, they are also fitted at this step. To derive the distance, the apparent dereddened mean $V$ magnitude of DM Cyg has to be known as the IPM gives the absolute mean $V$ brightness of the star. To determine the apparent dereddened mean magnitude of DM Cyg, the standard $V$ magnitude of the comparison stars and the interstellar absorption $A_V$ should have to be known. Unfortunately, no standard magnitudes of the comparison stars have been published. We estimate the average of the dereddened $V$ brightnesses of the two comparison stars to be 11.333 mag transformed from their Tycho $B_T$, and $V_T$ magnitudes \citep{hip} and using  $A_V = 3.14E(B-V)= 0.558$ interstellar absorption value according to the \cite{schlegel} maps. It has to be emphasized, however, that the uncertainties of the standard $V$ magnitudes of the comparison stars and the interstellar absorption value affect only the distance estimate. Neither any other parameter nor their detected variations depend on the choice of the dereddened $V$ magnitudes of the comparison stars.

The uncertainties of the estimates of the absolute physical parameters listed in Table~\ref{pp} correspond to the standard deviations of the results of running the IP code with 16 different settings. These settings are the combinations of: $i)$ $B$, $V$ and $I_C$, or $V$, $B-V$ and $V-I_C$ data are used; $ii)$ the $V_{rad}$ curve is defined by Liu's template or it is calculated from the $I_C$ light curve; $iii)$ 2 different weights of the initial $V_{rad}$ curve are used; $iv)$ 2 different values of the $\Delta A(V_{rad})/ \Delta V_{Amp}$ ratio valid for Blazhko variables are applied \citep[see further details in ][]{ipm,mw2}. However, as it can be seen from the data listed in Table~\ref{pp}, the true uncertainties of the mean global parameters are larger than the estimated errors of the solutions obtained for fixed mass and metallicity values. The true possible parameter space comprises the full range of the solutions for the possible mass and metallicity ranges. Nevertheless, we are focusing on the variation in the mean physical parameters and not on their absolute values, and it is shown in the rest of this section that the uncertainties of the mean global physical parameters have no effect on their variation during the Blazhko cycle.

Though the amplitude of the modulation of DM Cyg is only about one tenth of that of MW Lyr, we have applied the IPM to the light curves in 10 different phases of the modulation in order to decide whether there are any detectable changes in the mean global parameters of DM Cyg during its Blazhko period. The distance and the mass are fixed to their possible values given in Table~\ref{pp} assuming [Fe/H]$=0.0$ metallicity. The zero points of the magnitude scales are also fixed correspondingly to the solution obtained for the mean light curves. 

Fig.~\ref{minden} shows the results for the solutions using $\mathfrak{M}=0.50 \mathfrak{M}_{Sun}$, $d=1179$~pc; $\mathfrak{M}=0.55 \mathfrak{M}_{Sun}$, $d=1215$~pc and $\mathfrak{M}=0.60 \mathfrak{M}_{Sun}$, $d=1233$~pc mass and distance values. In the left-hand panels, the variations in the observed mean quantities are plotted: total pulsation amplitude in $V$ band, the variation of the pulsation period derived from the phase variation of the $f_0$ pulsation frequency and different pulsation averages of the magnitudes and colours. The right-hand panels show the quantities derived from the IP method for three possible  mass/distance combinations: the mean values of the radius, the surface gravity, the absolute visual brightness averaged by magnitude and intensity units, the luminosity and the effective temperature. The different combinations of the mass and distance give very similar results, the amplitudes and phases of the detected  changes in the mean physical parameters during the Blazhko cycle hardly change, only their averages over the Blazhko cycle are different, corresponding to their respective values given in Table~\ref{pp}. If atmosphere models with [Fe/H]=$-0.1$ are used, the results on the variations of the physical parameters during the Blazhko
cycle are not changed.

In order to get a real estimate of the uncertainties of the derived quantities in different phases of the modulation, the IP code has been run with 16 different settings for each data set again. 
Note that the zero points of the colour and magnitude scales defined by the dereddened magnitudes of the comparison stars, are also fixed now. These magnitudes were, however, allowed to vary i.e., they were also fitted when the mean global parameters were determined. This explains why the scatter of the results for the 16 different setting of the IP method is much smaller in the different phases of the modulation than the scatter of the results for the mean light curves.

The different intensity and magnitude averages of the $V$ light curve and the colour curves show hardly any variations, the amplitudes of their changes is only 0.002-0.004 mag. The $V$, and $B-V$ averages  show some small systematic variations of similar character like these averages in MW Lyr, but the $V-I$ averages show only scatter. In spite of that the observed mean magnitudes and colours vary only slightly during the Blazhko cycle in DM Cyg, by the aid of the IP method 0.3\% and 7 K systematic changes in the mean luminosity and temperature  of the star could be detected during its Blazhko cycle. The phase relation of these variations are the opposite, similarly as it was detected in MW Lyr. Both stars are the most luminous and the coolest at around the largest amplitude phase of the modulation.

The IP method calculates the mean physical parameters as their mathematical averages over the pulsation cycle. $F(\Phi) =\, <F(\varphi)>$, where $\Phi$ and $\varphi$ denote modulation and pulsation phases, respectively, and $< >$ denotes averaging by $\varphi$ for the whole pulsation cycle. For the luminosity, it means that the values plotted in Fig.~\ref{minden} correspond to:

$L(\Phi) = \,< L(\varphi)>\, =\, < 4 \pi \sigma R(\varphi)^2 \,\, T_{eff}(\varphi)^4>$,

\noindent as the IP method satisfies the Stephan-Boltzmann law in each phase of the pulsation ($\varphi$).

\cite{carney} showed from direct Baade-Wesselink analysis results that, even when the equilibrium luminosity and radius of pulsating variables equal with the arithmetic means of their variations over the pulsation cycle, the equilibrium temperature $T_{eq}$ is not the same as the mean temperature $<T_{eff}>$ of the star. The equilibrium temperature is defined then as 

$T_{eq}=(L_{eq}/4\pi\sigma R_{eq}^2)^{1/4}= (L(\Phi)/4\pi\sigma R(\Phi)^2)^{1/4} $ 

\hskip 20 pt $ = (<L(\varphi)>/4\pi\sigma <R(\varphi)>^2)^{1/4}$. 

\noindent Consequently, the relation between the arithmetic means of the physical parameters shown in Fig.~\ref{minden} deviates to some extent from the theoretical Stefan-Boltzmann law.

The $0.13\%$ variation in the mean radius is in good agreemant with the $0.17\%$ changes of the pulsation period (the pulsation equation requires $\Delta P/P \approx 3/2 \Delta R/R)$.  The pulsation period changes are determined from the phase differences of the $f_0$ pulsation frequency in different phases of the modulation, but when deriving the radius variation the IP method does not utilize this information in any way. Consequently, the determined period and  radius variations are completely independent quantities. The fact, that they show such a good agreement is a great support  both to the IP method, and  to our interpretation of the phase modulation as variations in the pulsation period.

Comparing the results obtained for DM Cyg with that of MW Lyr \citep[Fig. 14 in][]{mw2} the similar behaviour of the two stars are conspicuous. The only difference is in the detected amplitudes of the variations of the different parameters, each being about one tenth  of their detected amplitudes in MW Lyr, in accordance with the different amplitudes of the modulation of the two stars.

\section{The photographic and photoelectric data}

\begin{figure}
  \centering
  \includegraphics[width=6.9cm]{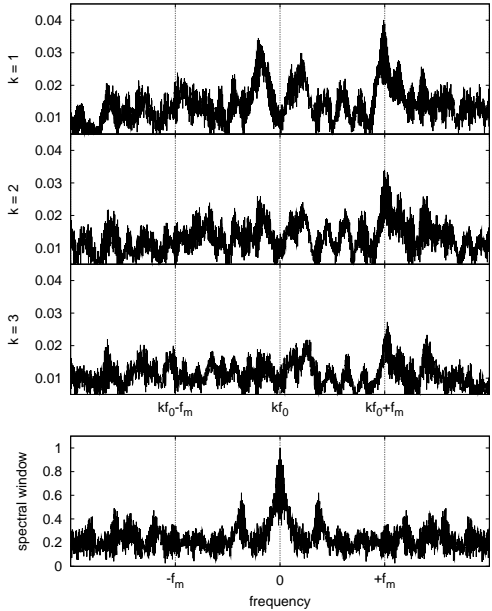}
  \caption{ Amplitude spectrum of the residual photographic light curve (after the removal of the pulsation signal) in the vicinity of the $f_0, 2f_0$ and $3f_0$  pulsation frequencies. The residuals show signals at around $f_0+f_m$, $2f_0+f_m$ and $3f_0+f_m$  frequencies with $0.02-0.03$ mag amplitude. The $f_0$ and $f_m$ frequencies detected in the photographic data are 2.3817564 and 0.0940, respectively. Bottom panel shows the spectral window of the photographic observations. }
  \label{pgsp}
\end{figure}

The photographic light curve of DM Cyg was shown in \cite{hurta}. Although the Konkoly photographic data are very sparse, the observations covered 24 years with a 13-year gap, we tried to analyse this data set also searching for any sign of light curve modulation. No indication of light curve modulation has been found in any of the two parts of the photographic data, but there is some hint that modulation frequency components appear at  $kf_0+f_m, (k=1,2,3)$ positions in the residual spectrum of the complete photographic data set (see Fig.~\ref{pgsp}). The highest modulation peaks are not exactly at the same separations for $f_0+f_m$, $2f_0+f_m$ and $3f_0+f_m$. The best solution which gives the smallest residual for the photographic data can be gained with 2.3817564 c/d (0.41985822 d) and 0.0940 c/d (10.636 d) pulsation and modulation frequencies, respectively.

\begin{figure}
  \centering
  \includegraphics[width=4.cm, angle=-90]{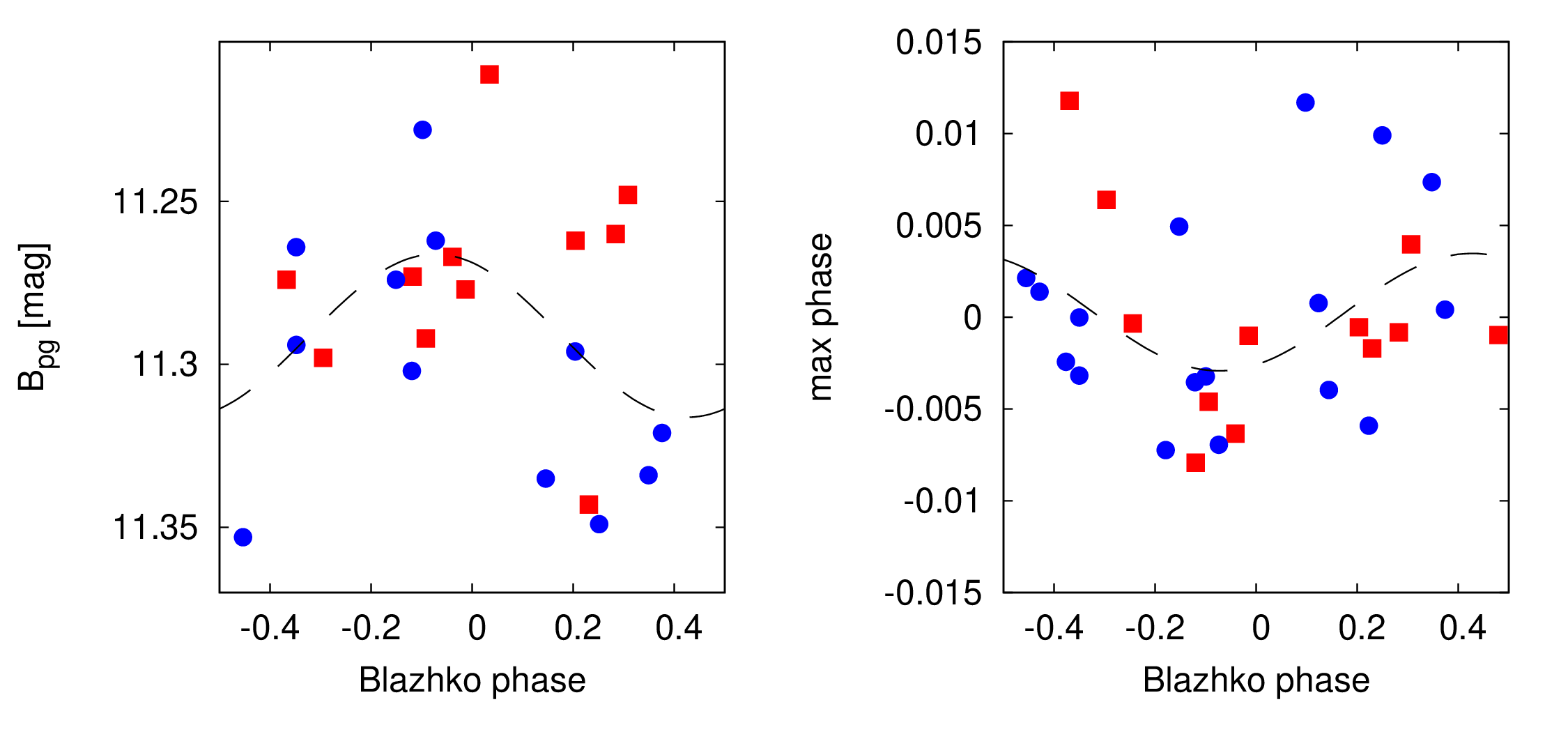}
  \caption{Maximum brightness and phase of the $B_{pg}=11.5$~mag brightness values on the rising branch  (0.2 mag below maximum brightness) for the photographic observations phased with the supposed modulation period (10.636 days) are shown in the left-hand and right-hand panels, respectively. Data from the two parts of the photographic observations are shown with different symbols. According to these plots, there is some indication that the maximum brightness and the phase of the rising branch of DM Cyg varied with small amplitude during the time of the photographic observations as well.}
  \label{maxpg}
\end{figure}

\begin{figure}
  \centering
  \includegraphics[width=5.6cm,angle=-90]{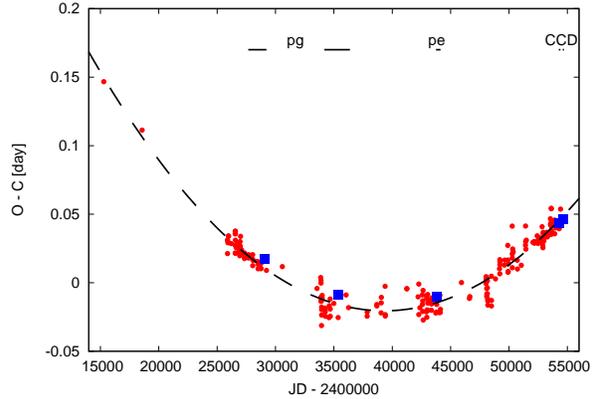}
  \caption{
The $O-C$ diagram of DM Cyg. Literature data are from the GEOS database (dots), the most deviant data are omitted. The Konkoly photographic, photoelectric and CCD normal maximum timings listed in Table~\ref{maxtime} are plotted by squares, the time intervals of the Konkoly observations are indicated by horizontal lines in the top of the figure. The $O-C$ indicates steady period increase during the $\sim$hundred years of the observations. }
  \label{oc}
\end{figure}

Fig.~\ref{maxpg} plots the maximum brightness and the phase of the rising branch at $B_{pg}=11.5$~mag brightness values phased with the supposed 10.636 d modulation period. The phase of a given magnitude on the rising branch can be determined with higher accuracy than the phase of the maximum, therefore we use this quantity to measure the amplitude of the phase modulation component in the photographic data. Data belonging to the two parts of the observations are denoted by different symbols in Fig.~\ref{maxpg}. According to these plots the amplitudes of the amplitude and phase modulations were not larger at the time of the photographic observations than today.

The photographic data were also analysed taking into account the period decrease of the pulsation that took place during the time interval of the observations, but the results were only marginally different from that obtained from the original photographic data.

The photoelectric observations comprise data only from four nights which does not allow to check the light curve modulation. The photoelectric data were therefore used only to determine one normal maximum timing listed in Table~\ref{maxtime}. 

The normal maximum timings of the Konkoly observations given in Table~\ref{maxtime} together with literature data collected in the GEOS database are plotted in Fig.~\ref{oc}. Based on the GEOS data \cite{geos} determined $\beta = 0.091 \mathrm{ d\,Myr^{-1}}$ period increase rate for DM Cyg. The addition of the Konkoly data does not modify this result.

From the analyses of the CCD and photographic data of DM Cyg we conclude that the period changes of the pulsation and modulation have the opposite direction, when the pulsation period of DM Cyg was shorter by 0.000005 d then the modulation period was 0.066 d longer than today. These values correspond to $\frac{dP_B}{dP_0}=-13200$ or $\frac{dP_B}{P_B} / \frac{dP_0}{P_0} = -523$ period change rates. However, these data have to be taken with caution partially because the uncertainties of the results derived from the photographic data and also because pulsation and modulation period changes of those Blazhko variables where the periods could be determined for several epochs show that there is no strict relation between the two periods, for some time intervals even the sign of the period change ratio can change \citep[e.g., in RV UMa][]{rvuma}. Therefore, the period change ratios defined by only two-epoch data should be missleading in some cases. We also have to note, however, that the complex period change behaviour of the modulation is usually connected to complex changes of the pulsation period and  of the properties of the modulation, too \citep[e.g., in RR Gem and XZ Cyg][]{rrgII,lacluyze}. Probably, the steady period change of the pulsation of DM Cyg is accompanied with steady period change of its modulation and with the stability of its modulation properties.

\section{Summary}

In the present paper the light curve modulation of DM Cyg, a fundamental mode RR Lyrae variable has been investigated. Most of the results are very similar to those found in the analysis of MW Lyrae, a large modulation amplitude Blazhko variable, but on a significantly smaller scale. The 0.07 mag amplitude of maximum brightness variation of DM Cyg is only 10\% of the modulation amplitude of MW Lyr, accordingly, all the changes detected in DM Cyg during its Blazhko period are about 10\% of the detected changes in MW Lyr. 

The phase and amplitude relations of the amplitude and phase modulation components are similar in DM Cyg and MW Lyr. The variations of the mean magnitudes and colours in DM Cyg where they can be detected are also similar to their counterparts in the MW Lyr data. Therefore, it is not surprising that the derived changes in the mean global physical parameters of DM Cyg resemble the changes found in the mean parameters of MW Lyr only on a much reduced scale. 

The only notable difference between the character of the modulations of the two stars is found when the changes of the Fourier parameters of the light curves during the Blazhko cycle is investigated. While in the case of MW Lyr the phases of $2f_0,$ and $3f_0$ change in line with the phase variation of $f_0$ (i.e, there is no significant changes detected in the $\varphi(f_{k1})$ phase differences), when analysing the light curves of DM Cyg  $90\degr$ phase differences between the phase variation of the $f_0$ pulsation frequency and the phase variations of the higher order pulsation components are found. In spite of this, the data transformation  that corrects the times of the observations taking into account the phase variation of the $f_0$ pulsation frequency which completely separated the amplitude and phase modulation components of the light curve modulation of MW Lyr seems to work also on the light curve of DM Cyg. 

The modulation behaviour of other Blazhko variables can differ, however, more significantly from the modulations of DM Cyg and MW Lyr. There are Blazhko stars, where the phase relation of the  amplitude and phase modulation components contrast with the phase relations of DM Cyg and MW Lyr. Also, the detected changes in the mean colours may vary differently e.g,  in SS Cnc \citep{ssc}. Therefore, we cannot draw major conclusions from the similarity of the results obtained for DM Cyg and MW Lyr before analysing other Blazhko variables of different character on a similar way.

\section{Acknowledgments}

The financial support of OTKA grants T-068626 and T-048961 is acknowledged. We wish to thank K. Ol\'ah for obtaining the photoelectric observations of DM Cyg. Zs.K. is a grantee of the Bolyai J\'anos Scholarship of the Hungarian Academy of Sciences.

\end{document}